\documentclass{PoS}
\usepackage{amsmath,amssymb,amsbsy,amsfonts,bm}
\usepackage[mathscr]{euscript}

\def\gtwid{{\,\raise.3ex\hbox{$>$\kern-.75em\lower1ex\hbox{$\sim$}}\,}}
\def\ltwid{{\,\raise.3ex\hbox{$<$\kern-.75em\lower1ex\hbox{$\sim$}}\,}}

\def\tr{\rm{tr}}
\def\chpt{\raise0.2ex\hbox{$\chi$}PT}
\def\bmchpt{\raise0.2ex\hbox{${\bm \chi}$}PT}
\def\schpt{S\raise0.2ex\hbox{$\chi$}PT}
\def\rschpt{rS\raise0.2ex\hbox{$\chi$}PT}
\def\rhmschpt{rHMS\raise0.2ex\hbox{$\chi$}PT}

\def\cH{{\cal H}}
\def\cI{{\cal I}}

\def\cO{{\cal O}}

\def\cQ{{\cal Q}}

\def\cT{{\cal T}}

\def\cW{{\cal W}}

\def\eqn#1{\label{eq:#1}}

\def\eq#1{Eq.~(\ref{eq:#1})}
\def\eqsthru#1#2{Eqs.~(\ref{eq:#1}) through (\ref{eq:#2})}

\def\figref#1{Fig.~\ref{fig:#1}}

\def\gtwid{{\,\raise.3ex\hbox{$>$\kern-.75em\lower1ex\hbox{$\sim$}}\,}}
\def\ltwid{{\,\raise.3ex\hbox{$<$\kern-.75em\lower1ex\hbox{$\sim$}}\,}}

\newcommand{\BE}{\begin{equation*}}
\def\EE{\end{equation*}}
\def\BEA{\begin{eqnarray*}}
\def\EEA{\end{eqnarray*}}

\def\BI{\begin{itemize}}
\def\EI{\end{itemize}}
\def\ED{\end{document}}

\title{Staggered Chiral Perturbation Theory for Neutral ${\bm B}$ Mixing}

\ShortTitle{Staggered Chiral Perturbation Theory for Neutral $B$ Mixing}

\author{\speaker{C.\ Bernard}\\
        Department of Physics, Washington University, St.~Louis, MO 63130, USA\\
        E-mail: \email{cb@wustl.edu}\\ }

\vspace{2mm}

\author{(The MILC Collaboration)}

\abstract{I describe a calculation of $B$ meson mixing at one-loop in
staggered chiral perturbation theory, for the complete set
of Standard Model and beyond-the-Standard Model operators.  The particular 
lattice representation of the continuum operators used
by the Fermilab Lattice/MILC collaborations (and earlier by
the HPQCD collaboration) turns out to be important, and results
in the presence of "wrong-spin" operators, whose contributions
however vanish in the continuum limit.   The relation between 
staggered and naive fermions also plays a key role.}

\FullConference{The 30$^{\rm th}$ International Symposium on Lattice Field Theory -- Lattice 2012\\
                June 24--29, 2012\\
                Cairns, Australia}

\begin{document}

\section{Introduction}

The mixing of neutral $B$ mesons provides a fertile area for 
tests of the Standard Model and 
sensitivity to new physics. In order to take advantage
of experimental measurements, lattice computations are as usual needed
to determine the hadronic matrix elements of the
effective weak operators. The mixing is dominated by the 
short distance contributions
from the operators  \cite{Gabbiani:1996hi}
\begin{eqnarray}
\cO_1 &=& (\bar b \gamma^\nu L q)\; [\bar b \gamma^\nu L q]\,, \qquad
\cO_2 = (\bar b L q) \;[\bar b L q]\,, \qquad
\cO_3 = (\bar b  L q]\; [\bar b  q) \,,\nonumber\\
\cO_4 &=& (\bar b L q)\; [\bar b  R q]\,,  \qquad
\cO_5 = (\bar b L q]\; [\bar b R q) \,,
\eqn{SUSY-basis}
\end{eqnarray}
where $\cO_1$ through $\cO_3$ control the mixing in the Standard Model, while
$\cO_4$ and $\cO_5$ can appear in beyond-the-Standard Model (BSM) theories.
Pairs of round or square parentheses in \eq{SUSY-basis} indicate how the color indices 
are contracted, and $R$ and $L$ are the right and left projectors.
Using Fierz transformations and parity, the mixing matrix element 
of any 4-quark operator with these quantum numbers can be written in terms of
the matrix elements of these five ``basis'' operators; see, for example,
Ref.~\cite{Bouchard-thesis} for details.

In a lattice computation, it is useful to be able to fit the lattice data
to a version of chiral perturbation theory that includes the effects of the
discretization errors associated with the lattice action.
In two  recent lattice calculations of
$B$ mixing \cite{Gamiz:2009ku,Bazavov:2012zs}, 
staggered light quarks are combined with non-staggered heavy quarks using
NRQCD \cite{NRQCD} or the Fermilab action \cite{El-Khadra:1996mp}.
In such cases, the appropriate chiral theory
is ``rooted, heavy-meson staggered chiral perturbation theory'' (\rhmschpt) \cite{Aubin:2005aq}.
Here, I describe a calculation of $B$ mixing to one-loop order in \rhmschpt. Roughly speaking, 
the calculation is to leading order in the heavy-quark expansion, although the largest
$1/m_B$ effects (the $B$-$B^*$ hyperfine splitting $\Delta^*$ and the $B_s$-$B_d$ flavor splitting) 
are also included. This is a systematic approximation in the power
counting introduced by Boyd and Grinstein \cite{Boyd:1994pa} and discussed recently 
in Ref.~\cite{Bazavov:2011aa} for the
lattice calculation of heavy-light meson decay constants.

The \rhmschpt\ calculation needs to take into account the form of the 
lattice operator used to approximate the continuum one.  
References~\cite{Gamiz:2009ku,Bazavov:2012zs} construct
the 4-quark operators as the local product of two local bilinears, each 
formed from a heavy antiquark field and a naive light-quark field.
Both the use of naive fields and the local nature of the 4-quark operator 
influence the form of the corrections at one-loop.

\section{Detailed Structure of the Operators}

Heavy-light bilinears and 4-quark operators are made by converting a one-component
staggered fermion $\chi(x)$ into a naive fermion 
$\Psi(x)$ following Ref.~\cite{Wingate:2002fh}, and then coupling it locally to the heavy-quark
field $Q(x)$:
\begin{eqnarray}
	 &\operatorname{bilinear:} &\quad\bar Q(x)\,\Gamma\, \Psi(x)\ , \eqn{bilinear} \\
	 &\operatorname{4-quark\ operator:} &\quad \bar Q(x)\,\Gamma\, \Psi(x)\; \bar Q(x)\,\Gamma'\, \Psi(x) \eqn{4qop} \ ,
\end{eqnarray}
where $\Gamma$ and $\Gamma'$ are Dirac spin matrices. As we will see,
bilinears of this form work exactly as desired in the lattice simulation, but the
local product of the bilinears in the 4-quark operators introduces contributions
from operators with wrong taste and spin ({\it i.e.}, spin different from  $\Gamma \otimes \Gamma'$).

The naive light-quark action can be written as four copies of the staggered
action:
\begin{equation}
\eqn{omega}
\Psi(x) = \Omega(x)\;\underline{\chi}\;, \qquad \Omega(x) = 
\gamma_0^{x_0}\,\gamma_1^{x_1}\,\gamma_2^{x_2}\,\gamma_3^{x_3}\ .
\end{equation}
Here $\underline{\chi}$ is a ``copied'' staggered
field, with each Dirac component $\underline{\chi}{}_i $ separately having the staggered
action.  I will call the $SU(4)$ symmetry that acts on index $i$ ``copy symmetry.''
Copy symmetry is an exact lattice symmetry, so copied and uncopied propagators are related by
\begin{equation}
\eqn{copied-prop}
\langle \underline{\chi}{}_i(x)\, \underline{\bar\chi}{}_{i'}(y)\rangle = \delta_{i,i'}\;
\langle\chi(x)\, \bar\chi(y)\rangle \ .
\end{equation}
where $\chi$ is the normal (uncopied) staggered field.  
An interpolating field $\cH(x)$ for a heavy-light pseudoscalar meson is
\begin{equation}
\eqn{interp-field}
\cH(x) = \bar Q(x)\,\gamma_5\, \Psi(x) = \bar Q(x)\,\gamma_5\, \Omega(x)\,\underline{\chi}(x)\ .
\end{equation}
In the simulations, $\cH(x)$ is always summed over a time-slice (either explicitly, or implicitly by using translation invariance).  
To leading order in $a$, $Q(x)$ varies smoothly
between neighboring spatial sites (up to gauge transformation), 
but $\underline{\chi}$ does not, due to taste doubling.
Staggered fields are, however, smooth in the spin-taste basis on the doubled lattice, so we 
need to sum the field within a hypercube to expose the structure of the operators. 

I focus first on the average of $\cH(x)$ over a spatial cube.  Let
$x=(t,{\bm x})$ with ${\bm x}=2{\bm y}$ even, 
and let $\eta=(\eta_0,{\bm \eta})$ be a 4-vector with
all components 0 or 1. 
For $t$ even ($t=2\tau$), 
\begin{eqnarray}
&&\cH^{\rm(av)}(t,{\bm x}) =  \frac{1}{8} \sum_{\bm \eta}\bar Q(t,{\bm x}+{\bm \eta})\,\gamma_5\, 
   \Omega(2\tau,{\bm \eta})\,\underline{\chi}(2\tau,2{\bm y}+{\bm \eta}) 
\cong\frac{1}{8}\; \bar Q(t,{\bm x})\,\gamma_5  \sum_{\bm \eta} 
            \Omega({\bm \eta})\,\underline{\chi}(2\tau,2{\bm y}+{\bm \eta}) \eqn{H-av0} \nonumber \\
&&\hspace{10mm}\cong \frac{1}{16}\; \bar Q(t,{\bm x})\,\gamma_5\; \sum_{\eta}
\big[\Omega(\eta)\,\underline{\chi}(2\tau+\eta_0,2{\bm y}+{\bm \eta})+
	(-1)^{\eta_0}\,\Omega(\eta)\,
\underline{\chi}(2\tau+\eta_0,2{\bm y}+{\bm \eta}) \big] \ . 
\end{eqnarray}
Inserted gauge links for point-split quantities are implicit.
For $t$ odd, the result in \eq{H-av0} is the same except the last term
changes sign.  This is the usual oscillating state with opposite parity.

For simplicity, I assume from now on that the oscillating  state is removed by 
the fitting procedure, and that  all components of $x$ are even.
Then
\begin{equation}
\cH^{\rm(av)}(x) \to  \frac{1}{16}\; \bar Q(x)\,\gamma_5\, \sum_{\eta}
\Omega(\eta)\,\underline{\chi}(2y+\eta)\ . \eqn{H-av1}
\end{equation}
To convert to a spin-taste basis, one can use \cite{TASTE-REPRESENTATION}
\begin{equation}
q^{\alpha a}_i(y) =  \frac{1}{8}\; \sum_{\eta}
\Omega^{\alpha a}(\eta)\,\underline{\chi}{}_i(2y+\eta)\ , \eqn{CopyKluberg} 
\end{equation}
where $\alpha$ is a spin index, $a$ is a taste index, and $i$ is a (trivially inserted) copy
index.
This implies that taste and copy indices are coupled in $\cH$. We get (with spin indices implicit from now on)
\begin{equation}
\cH^{\rm(av)}(x) \to  \frac{1}{2}\; \bar Q(x)\,\gamma_5\, q^a_i(y)\, \delta^a_i\ ,
\eqn{H-av2}
\end{equation}
where the arrow signifies that oscillating states and $\cO(a)$ corrections are being dropped.
Using \eq{copied-prop}, we see that the
contraction of $\cH$ with $\cH^\dagger$ is automatically averaged  over tastes:
\begin{equation}
\langle \cH(x)\,\cH^\dagger(x')\rangle \sim  \frac{1}{4}\; \langle\bar Q(x)\gamma_5 q^a(y)\; 
\bar q^a(y')\gamma_5 Q(x')\rangle \ .
\eqn{H-prop}
\end{equation}
Note that the interpolating field constructed from the naive light quark
gives us the desired pseudoscalar spin in the spin-taste basis (aside from the dropped oscillating state).

I now turn to the 4-quark operators constructed as in \eq{4qop}.  The two bilinears are
not separately summed over space, but they can be disentangled using 
\begin{equation}
\frac{1}{256}\sum_K tr\!\left(\Omega(\eta)\,K\, \Omega^\dagger(\eta)\, K\right)\; 
tr\!\left(\Omega(\eta')\,K\, \Omega^\dagger(\eta')\, K\right)
=   \delta_{\eta,\eta'}\ ,  \eqn{eta-ident}
\end{equation}
where $K$  runs over the 16 independent Hermitian gamma matrices. Following the same
kinds of manipulations as in \eqsthru{interp-field}{H-av2}, one then finds 
\begin{eqnarray}
\cO^{(av)}_n(x) 
&\to& \frac{1}{4} \sum_K (\bar Q  \Gamma_n K q^c_k \;
\bar Q  \Gamma'_n K q^d_\ell)\; K_{ck} K_{d\ell}\ , \eqn{O-av1}
\end{eqnarray}
where $\Gamma_n$ and $\Gamma'_n$ are the Dirac spin matrices of the operators.
Here the contributions with $K\not=I$ clearly have the wrong (undesired) spin, as well as 
non-trivial coupling of taste ($c,d$) and copy ($k,\ell$) indices.  
Putting \eq{H-av2} together with \eq{O-av1} and using
 copy symmetry gives
\begin{equation}
\langle \cH^\dagger \cO^{(av)}_n \cH^\dagger \rangle \propto \langle D^{ac} D^{ed}  
\rangle  K_{ca} K_{de} + \textrm{(second equivalent contraction)}\ ,
\end{equation}
where $D^{ac}$ is the light-quark propagator (in a given background) 
for taste $a$ into taste $c$.
When taste symmetry is exact, $\langle D^{ac} D^{ed}  \rangle\propto \delta_{ac} 
\delta_{ed}$, so $K=I$, and only the correct spins contribute.  This
shows that 4-quark operators constructed as in \eq{4qop} will have the correct (desired) spin
in the continuum limit, as well as at tree level in \rhmschpt, which respects taste symmetry.
But, at one loop (and $a\not=0$), taste-violations imply that $\langle D^{ac} D^{ed}  \rangle$ 
is not 
necessarily proportional to $\delta_{ac} \delta_{ed}$. Thus one-loop diagrams will have 
contributions from wrong-spin operators.

\section{Calculation of the \chpt\ Diagrams at One Loop}

\begin{figure}
\begin{center}
\begin{tabular}{c c}
\includegraphics[width=5.6cm]{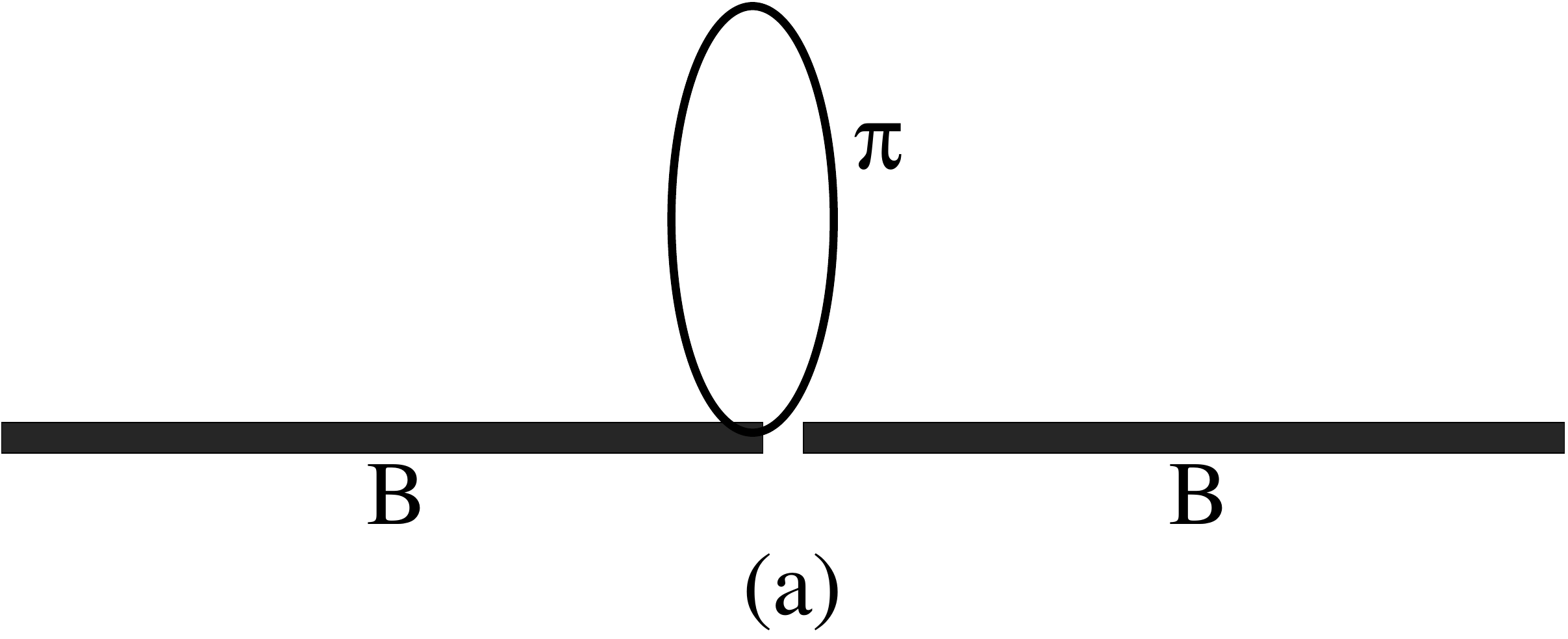}
&
\includegraphics[width=5.6cm]{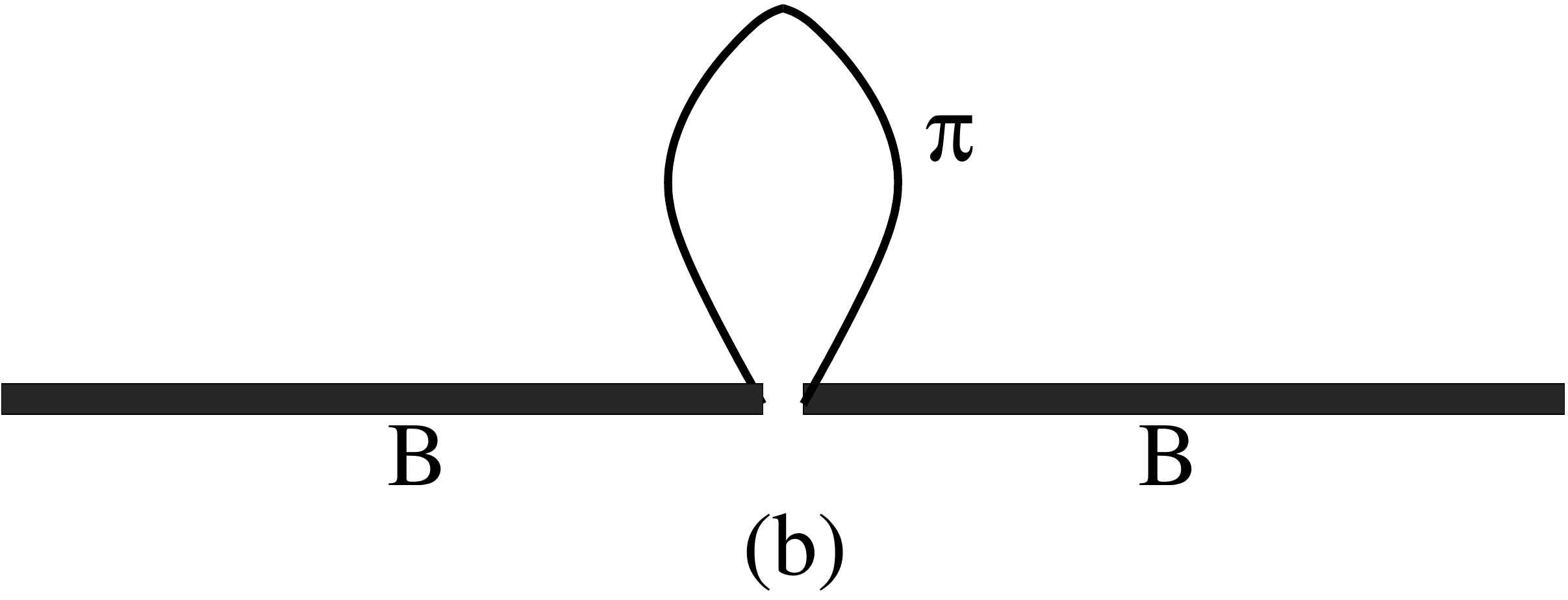}
\end{tabular}
\includegraphics[width=6.2cm]{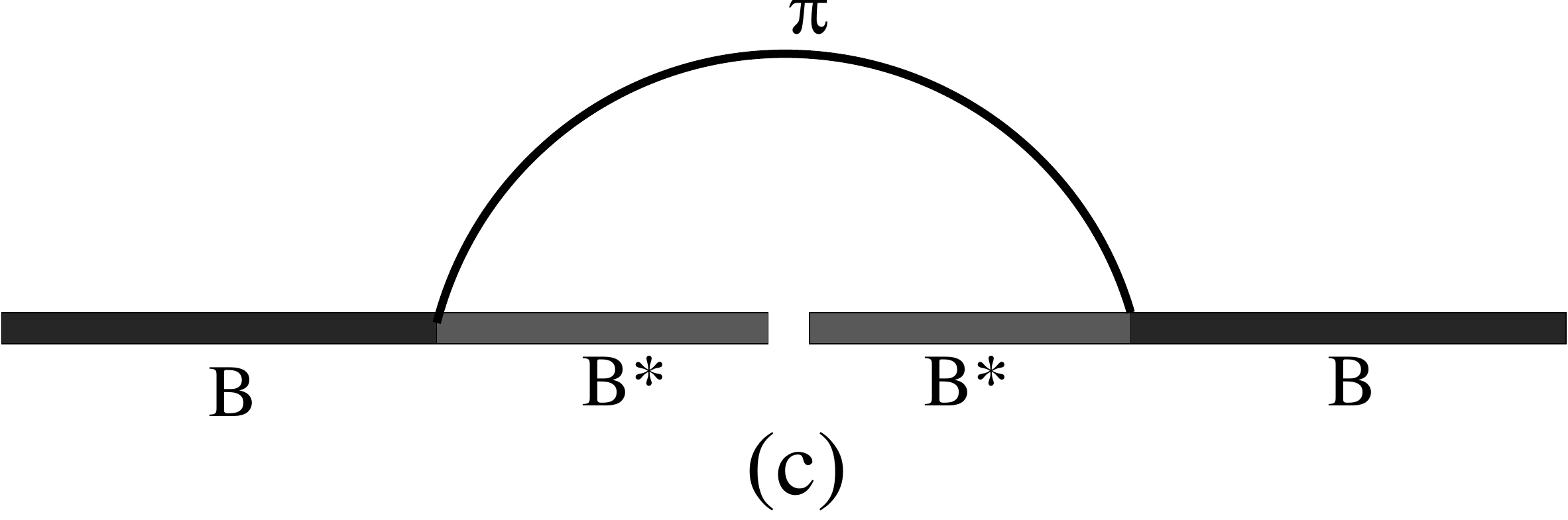}
\end{center}
\vspace{-7mm}
\caption{Meson-level chiral diagrams for the mixing matrix elements. The gap between
the two $B$ (or two $B^*$) fields represents the insertion of the 4-quark operator.
Diagrams (a) and (b) are tadpole diagrams, which are distinguished by whether the
contracting pions come from the same meson field (loosely speaking, the same bilinear)
or from different meson fields. Diagram (c) is the sunset diagram.}
\label{fig:meson-diagrams}
\end{figure}

We now are ready to perform the calculation of the one-loop chiral corrections to matrix
elements of the 4-quark operators. Aside from the wave-function  renormalization diagram,
which does not involve the operators and can be taken unchanged from the literature 
\cite{Aubin:2005aq,Aubin:2007mc}, there are three diagrams, shown in \figref{meson-diagrams}.

\begin{figure}
\begin{center}
\begin{tabular}{c c}
\includegraphics[width=6.8cm]{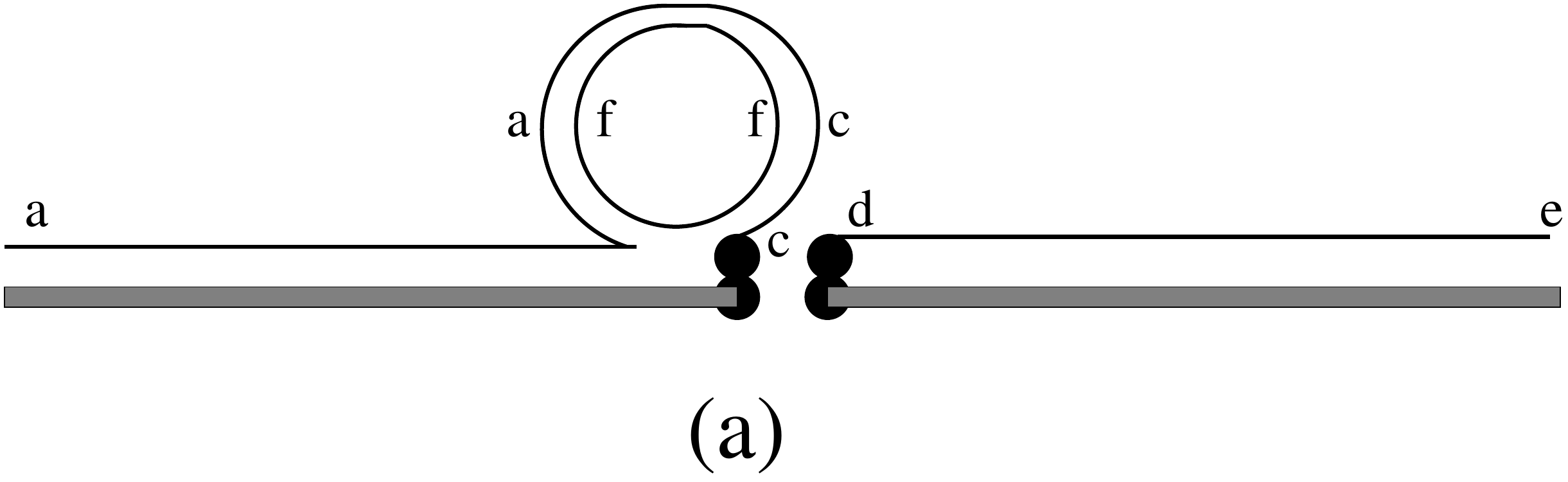}
&
\includegraphics[width=6.8cm]{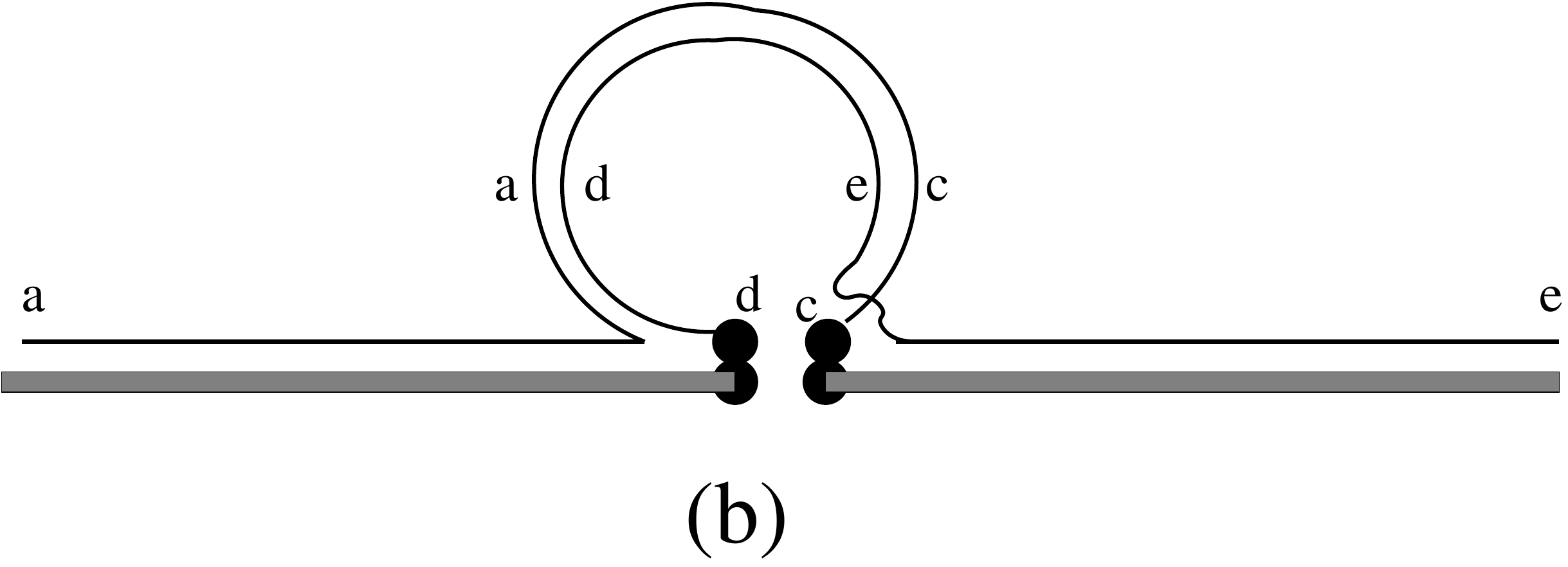}
\end{tabular}
\end{center}
\vspace{-7mm}
\caption{Examples of quark-flow tadpole diagrams.  The filled, black circles 
represent the heavy and light quarks in the 4-quark operator.
Diagram (a) contributes to the meson
diagram \protect\figref{meson-diagrams}(a), and diagram (b) contributes to
diagram \protect\figref{meson-diagrams}(b). Indices a--f label tastes.}
\label{fig:quark-diagrams}
\end{figure}

Because copy and taste indices are coupled in \eq{O-av1}, and because copy indices follow
the quark contractions (\eq{copied-prop}), we need to consider the quark flow of the meson
diagrams in order to compute them.  As an example, \figref{quark-diagrams} shows 
a possible quark flow for each of the tadpole meson diagrams.  These particular
quark flows have
``connected'' pion propagators without hairpin vertices; other tadpole flows with 
either taste-singlet (physical) or taste-violating hairpin contributions are also possible.

Labeling the taste of the pion in the loop by $\Xi$  ($\Xi=1,\dots,16$),
diagram \figref{quark-diagrams}(a) is proportional to
$\Xi_{af}\Xi_{fc}\delta_{ed}
K_{ca}K_{de} =  [\tr(K)]^2$, where the factors of $K$ come from \eq{O-av1}. In this
case, only the correct spin contributes ($K=I$). 
On the other hand,
diagram \figref{quark-diagrams}(b) is proportional to
$\Xi_{ad}\;\Xi_{ec} K_{ca}\,K_{de} = \tr(\Xi\; K\; \Xi\; K)$.
If the propagator were independent of $\Xi$, the sum on tastes would again give $K=I$.
Due to taste violations, however, the pion propagator 
depends on whether $\Xi$ is P, A, T, V, or I.  This implies
that various $K$ values are possible, giving wrong-spin contributions.

The fact that wrong spins enter means that other operators appear, and they in turn have
different chiral representatives than the original operator does.
Fortunately, the basis of $\cO_1,\dots,\cO_5$ is complete, and
the chiral representatives of all these
operators are given by Detmold and Lin \cite{Detmold:2006gh}.  Generalizing to operators with
light-quark tastes $c$, $d$, we have
\begin{eqnarray}
O^{xc;xd}_{1} &=&  \beta_{1} \left [ 
            \Big( \sigma P^{(b)\dagger} \Big)_{x,c}
            \Big( \sigma P^{(\bar{b})} \Big)_{x,d}
          + \Big( \sigma P^{\ast (b)\dagger}_{\mu} \Big)_{x,c}
            \Big( \sigma P^{\ast (\bar{b}),\mu} \Big)_{x,d}
                 \right ]  \hspace{3mm}[{\bf or\ } c\leftrightarrow d],
\\
 O^{xc;xd}_{2(3)} &=&  \beta_{2(3)} 
            \Big( \sigma P^{(b)\dagger} \Big)_{x,c}
            \Big( \sigma P^{(\bar{b})} \Big)_{x,d} 
            +  \beta_{2(3)}^\prime
            \Big( \sigma P^{\ast (b)\dagger}_{\mu} \Big)_{x,c}
            \Big( \sigma P^{\ast (\bar{b}),\mu} \Big)_{x,d}
\hspace{3mm}[{\bf or\ } c\leftrightarrow d], \eqn{chiral-ops2}
\\
 O^{xc;xd}_{4(5)} &=&  \frac{\beta_{4(5)}}{2}  \left [
            \Big( \sigma P^{(b)\dagger} \Big)_{x,c}
            \Big( \sigma^{\dagger} P^{(\bar{b})} \Big)_{x,d}  +
            \Big( \sigma^{\dagger} P^{(b)\dagger} \Big)_{x,c}
            \Big( \sigma P^{(\bar{b})} \Big)_{x,d} \right ]
\nonumber \\
&&\hspace{-7mm}           + \frac{\beta_{4(5)}^\prime}{2} \left [
            \Big( \sigma P^{\ast (b)\dagger}_{\mu} \Big)_{x,c}
            \Big( \sigma^{\dagger} P^{\ast (\bar{b}),\mu} \Big)_{x,d} +
            \Big( \sigma^{\dagger} P^{\ast (b)\dagger}_{\mu} \Big)_{x,c}
            \Big( \sigma P^{\ast (\bar{b}),\mu} \Big)_{x,d}\right ]\hspace{3mm}[{\bf or\ } c\leftrightarrow d].
\end{eqnarray}
where $P$ and $P^*$ are heavy-light meson fields, $\sigma$ is the pion field, and 
$x$ is the light 
flavor. The
effect of copy indices is to enforce the contraction of the external quark of taste $a$
with the quark in the operator of taste $c$, and similarly for $e$ and $d$.

We then write the matrix element for operator $\cO_n$ as
\begin{equation}
\langle \overline{B}_x^0|\cO_n^x|B_x^0 \rangle =
\beta_n \left(1+
{\cal
W}_{B}+
{\cal T}^{(n)}_x + \tilde{\cal T}^{(n)}_x\right)
+ \beta'_n\left({\cal
Q}^{(n)}_x+ \tilde{\cal Q}^{(n)}_x\right)+ \textrm{analytic terms}, 
\eqn{O2-5tot}
\end{equation}
where $\cW_B$ is the $B$ wave-function renormalization, $\cT$ and $\tilde \cT$ 
are the right- and wrong-spin tadpole diagrams, and 
$\cQ$ and $\tilde \cQ$ are the right- and wrong-spin sunset diagrams.  
In the special case of operator $\cO_1$, 
 $\beta'_1= \beta_1$ by a heavy-quark spin argument \cite{Detmold:2006gh}.

The diagrams for operator $\cO_1^x$ then give
\begin{eqnarray}
 {\cal T}_{x}^{(1)} &\hspace{-4mm}=\hspace{-4mm}&
\frac{-i}{f_\pi^2}\Bigg\{\frac{1}{16}\sum_{\mathscr{S},\rho}
N_\rho\,{\cal I}_{x\mathscr{S},\rho}
+\frac{1}{16}\sum_{\rho}N_\rho\,{\cal I}_{X,\rho}
+\frac{2}{3}\bigg[R^{[2,2]}_{X_I}\big(\{M^{(5)}_{X_I}\}
;\{\mu_I\}\big)\; \frac{\partial{\cal I}_{X,I}}{\partial
m^2_{X_I}}  \nonumber
\\ && -\hspace{-3mm}\sum_{j \in
\{M^{(5)}_I\}}D^{[2,2]}_{j,X_I}\big(\{M^{(5)}_{X_I}\};\{\mu_I\}\big){\cal
I}_{j,I} \bigg]
+a^2\delta'_{V}\bigg[R^{[3,2]}_{X_V}\big(\{M^{(7)}_{X_V}\}
;\{\mu_V\}\big)\; \frac{\partial{\cal I}_{X,V}}{\partial
m^2_{X_V}}   \nonumber \\ 
&& -\hspace{-3mm}\sum_{j \in
\{M^{(7)}_V\}}D^{[3,2]}_{j,X_V}\big(\{M^{(7)}_{X_V}\};\{\mu_V\}\big){\cal
I}_{j,V}
 \bigg] +\big(V\rightarrow A\big)\Bigg\}\ , \\
 \tilde {\cal T}_{x}^{(1)} &\hspace{-4mm}=\hspace{-4mm}&
\frac{-i}{f_\pi^2}\Bigg\{\frac{1}{16}
\bigg(-5{\cal I}_{X,P}-4{\cal I}_{X,A}+18{\cal I}_{X,T}-4{\cal I}_{X,V}-5{\cal I}_{X,I}\bigg) 
+\frac{2(\beta_2+\beta_3)}{\beta_1}\bigg({\cal I}_{X,A}-{\cal I}_{X,V}\nonumber\\
&&\hspace{5mm}+ a^2\delta'_{V}\Big[R^{[3,2]}_{X_V}\big(\{M^{(7)}_{X_V}\}
;\{\mu_V\}\big)\; \frac{\partial{\cal I}_{X,V}}{\partial
m^2_{X_V}}\hspace{3mm}
-\hspace{-3mm}\sum_{j \in
\{M^{(7)}_V\}}\hspace{-2mm}
D^{[3,2]}_{j,X_V}\big(\{M^{(7)}_{X_V}\};\{\mu_V\}\big){\cal
I}_{j,V}
 \Big]
\nonumber \\ 
&&\hspace{5mm}-a^2\delta'_{A}\Big[R^{[3,2]}_{X_A}\big(\{M^{(7)}_{X_A}\}
;\{\mu_A\}\big)\; \frac{\partial{\cal I}_{X,A}}{\partial
m^2_{X_A}} \hspace{3mm}
-\hspace{-3mm}\sum_{j \in
\{M^{(7)}_A\}}\hspace{-2mm}D^{[3,2]}_{j,X_A}\big(\{M^{(7)}_{X_A}\};\{\mu_A\}\big){\cal
I}_{j,A}
 \Big]
\bigg)\Bigg\}  \ , \\
{\cal Q}^{(1)}_{x} &=&
\frac{-ig_{B^*B\pi}^2}{f_\pi^2}\Bigg\{\frac{1}{16}\sum_{ \rho}
N_\rho\,{\cal H}^{\Delta^*}_{X,\rho}
+\frac{1}{3}\bigg[R^{[2,2]}_{X_I}\big(\{M^{(5)}_{X_I}\}
;\{\mu_I\}\big)\; \frac{\partial{{\cal
H}^{\Delta^*}_{X,I}}}{\partial m^2_{X_I}}  \nonumber
\\ &&\hspace{10mm} -\hspace{-3mm}\sum_{j \in
\{M^{(5)}_I\}}D^{[2,2]}_{j,X_I}\big(\{M^{(5)}_{X_I}\};\{\mu_I\}\big){\cal
H}^{\Delta^*}_{j,I} \bigg] \Bigg\}\ ,  \\
\tilde{\cal Q}^{(1)}_{x} &=&
\frac{-ig_{B^*B\pi}^2}{f_\pi^2}\Bigg\{
\frac{1}{16}
\bigg(-5{\cal H}^{\Delta^*}_{X,P}-4{\cal H}^{\Delta^*}_{X,A}+18{\cal H}^{\Delta^*}_{X,T}-4{\cal H}^{\Delta^*}_{X,V}-5{\cal H}^{\Delta^*}_{X,I}\bigg) 
+\frac{2(\beta'_2+\beta'_3)}{\beta_1}\Big({\cal H}^{\Delta^*}_{X,A}\nonumber\\
&&\hspace{5mm}-{\cal H}^{\Delta^*}_{X,V} 
+ a^2\delta'_{V}\Big[R^{[3,2]}_{X_V}\big(\{M^{(7)}_{X_V}\}
;\{\mu_V\}\big)\; \frac{\partial{\cal H}^{\Delta^*}_{X,V}}{\partial
m^2_{X_V}}   
-\hspace{-3mm}\sum_{j \in
\{M^{(7)}_V\}}\hspace{-3mm}D^{[3,2]}_{j,X_V}\big(\{M^{(7)}_{X_V}\};\{\mu_V\}\big){\cal
H}^{\Delta^*}_{j,V}
 \Big]
\nonumber \\ 
&&\hspace{5mm}-a^2\delta'_{A}\Big[R^{[3,2]}_{X_A}\big(\{M^{(7)}_{X_A}\}
;\{\mu_A\}\big)\; \frac{\partial{\cal H}^{\Delta^*}_{X,A}}{\partial
m^2_{X_A}} 
-\hspace{-3mm}\sum_{j \in
\{M^{(7)}_A\}}\hspace{-4mm}D^{[3,2]}_{j,X_A}\big(\{M^{(7)}_{X_A}\};\{\mu_A\}\big){\cal H}^{\Delta^*}
_{j,A}
 \Big]
\bigg)\Bigg\} \ .
\end{eqnarray}
The chiral logarithm functions $\cI$ and $\cH^{\Delta^*}$ are defined in Ref~\cite{Detmold:2006gh},
the subscripts on these functions give the flavor and taste of the relevant meson, $\mathscr{S}$ runs
over the sea quarks, $\rho$ sums over taste representations $P,A,T,V,I$ (with degeneracies $N_\rho$),
$X$ is the $x\bar x$ meson,
and other notation is given in Refs.~\cite{Aubin:2005aq,Aubin:2007mc}.
The contributions to operators $\cO_2$--$\cO_5$ have similar forms.

\section{Discussion}

Wrong spin/taste operators arise from the local operator construction and are $\cO(1)$ in
the lattice spacing.  Their contributions to matrix elements are suppressed to NLO because 
taste-symmetry violation is required.
This is different from mixings due to perturbative corrections, which are suppressed
by $\alpha_S/4\pi$. One-loop in \chpt\ then makes them
$\cO(a^2 \alpha_S/4\pi )$, which is effectively NNLO. So non-analytic chiral logarithms do not 
arise at
NLO from the perturbative corrections.
A similar statement is true  
for possible $\cO(a)$ and higher terms in the relation between the naive
quarks and the spin-taste basis.

The 
wrong-spin effects induce mixing in the \chpt\ of various operators.  However,
as long as all five $\cO_n$ are analyzed simultaneously,   
there are no new low-energy constants induced by these effects: the $\beta_n$ and $\beta'_n$ are
all already present in the continuum.

References \cite{Gamiz:2009ku,Bazavov:2012zs} focused on the calculation of the quantity $\xi 
\equiv (f_{B_s}\sqrt{\hat B_{B_s}})/( f_{B_d}\sqrt{\hat B_{B_d}})$, which comes from the 
matrix element of operator $\cO_1$.  
The chiral effects of the wrong spins were not known at the time
of the HPQCD calculation \cite{Gamiz:2009ku} and were therefore omitted from the analysis and error estimate.
In the Fermilab/MILC calculation \cite{Bazavov:2012zs}, the full \chpt\ expression 
was available, but the
matrix elements of operators other than $\cO_1$ were not calculated, preventing a direct
inclusion of the wrong-spin effects.  However, it was possible to estimate the error
of omitting these effects by using
a small subset of new data to investigate the other matrix elements.
The result, $\xi = 1.268(63)$  included a 3.2\% error from this effect, which was
the second largest source of error.
In the ongoing second-generation Fermilab/MILC project \cite{FreelandLAT12}, matrix elements
of all five operators $\cO_n$ are being computed, which means that the full
\chpt\ expression  can be used in the analysis, and there will be no ``wrong-spin
error.'' Of course, a chiral/continuum extrapolation error will remain.

I thank J.\ Laiho, R.S.\ Van de Water, and C.\ Bouchard for discussions and for
help with various aspects of the calculation.
This work has been partially supported by the Department of Energy, under grant number DE-FG02-91ER40628.

\end{document}